\documentclass[fleqn,usenatbib,useAMS]{mnras}

\usepackage{graphicx}	
\usepackage{wrapfig}
\usepackage{lscape}
\usepackage{rotating}
\usepackage{amsmath}	
\usepackage{amssymb}	
\usepackage{multicol}        
\usepackage{bm}		
\usepackage{pdflscape}	
\usepackage{natbib}
\usepackage{hyperref}
\usepackage{multirow}
\usepackage{arydshln}
\usepackage{float}
\usepackage{subfigure}

\newcommand{\kms}{\,km\,s$^{-1}$} 

\usepackage[T1]{fontenc}
\usepackage{ae,aecompl}

\title[]{Spectroscopic and geometrical evolution of the ejecta of the classical nova ASASSN-18fv}
 \author[Pavana et al.]{
 M. Pavana$^{1,2}$\thanks{E-mail: pavana@iiap.res.in, murali.pavana@gmail.com},
 A. Raj$^{1,3}$,
 T. Bohlsen$^{4}$,
 G. C. Anupama$^{1}$,
 Ranjan Gupta$^{5}$
 \newauthor
 and G. Selvakumar$^{1}$
 \\
 $^{1}$Indian Institute of Astrophysics, Kormangala Block II, Bengaluru 560034, Karnataka, India.\\
 $^{2}$Pondicherry University, R.V. Nagar, Kalapet, 605014 Puducherry, India.\\
 $^{3}$Department of Physics and Astrophysics, University of Delhi, Delhi 110007, India\\
 $^{4}$Mirranook Observatory, Booroolong Rd., Armidale, NSW, Australia.\\
 $^{5}$IUCAA, Post Bag 4, Ganeshkhind, Pune-411007, India.\\
 }

\date{Last updated XXX ; in original form YYY}
\pubyear{2019}

\begin{document}
\label{firstpage}
\pagerange{\pageref{firstpage}--\pageref{lastpage}}
\maketitle

\begin{abstract}
The optical spectroscopic observations of ASASSN-18fv observed from 2018 March 24 to 2019 Jan 26 are presented. The optical spectra are obtained from Mirranook, Vainu Bappu and  South African Astronomical observatories. The spectra are dominated by hydrogen Balmer, Fe II and O I lines with P-Cygni profiles in the early phase, typical of an Fe II class nova. The spectra show He I lines along with H I and O I emission lines in the decline phase placing the nova in the hybrid class of novae. The spectra show rapid development in high ionization lines in this phase. Analysis of the light curve indicate t$_2$ and t$_3$ values of about 50 and 70 days respectively placing the nova in the category of moderately fast nova. The ejecta geometry, inclination and position angle are estimated using morpho-kinematic analysis. The geometry of the ejecta is found to be an asymmetric bipolar structure with an inclination angle of about 53$^{\circ}$. The ejected mass using photo-ionization analysis is found to be 6.07 $\times$ 10$^{-4}$ M$_{\odot}$.
\end{abstract}





\section{Introduction}
ASASSN-18fv (V906 Car or Nova Car 2018) was discovered on 2018 March 16.32 UT by \cite{sta18} at magnitude of V = 10.4. A pre-discovery image obtained on 2018 March 16.227 UT indicated a magnitude of V = 10.21 $\pm$ 0.05 \citep{cor18}. \cite{str18} reported early optical spectroscopic observations of 2018 March 21 showing H$\alpha$, Ca II triplet, O I, Mg I, and several Fe II lines with P-Cygni profiles. The spectrum showed higher order Balmer and Paschen lines in absorption as well as a bump-like feature between 8300--9300 \AA\, with no obvious broad molecular features. They suggested the object to be the outburst of a young stellar object, or plausibly a stellar merger, although, a classical nova outburst could not be ruled out. \cite{luc18}, on the other hand, based on the presence of optically thick Fe II features in an optical spectrum obtained at a similar epoch, concluded the object to be a classical nova in the iron curtain phase. \cite{izz18} reported high resolution optical spectroscopic observations obtained on 2018 March 22. The continuum indicated an A-type main sequence star peaking around 5300 \AA. The H$\alpha$ line showed P-Cygni profile having blue-shifted heliocentric velocity, v $\sim$ -340 \kms, and a narrow emission line characterized by a measured FWZI = 670 km s$^{-1}$ and full width half maximum (FWHM) = 325 \kms. O I 7773 \AA\ line was found to be the strongest non-Balmer emission in the spectrum. \cite{izz18} also reported the presence of several Fe II multiplets with P-Cygni absorption centered at a heliocentric velocity of v $\sim$ -250 \kms\ as well as weak blue continuum and low expansion velocities and suggested the event to be of a different nature than a typical classical nova, more likely a red nova or a Helium flash explosion. However, \cite{rab18} based on a near-infrared (NIR) spectrum concluded that ASASSN-18fv belongs to the hybrid class of novae. The spectrum (range 0.8--1.7 $\mu$m, R$\sim$3500) obtained on 2018 April 1.23 UT showed a large number of emission lines of H I, He I, Ca II, N I, O I and C I with relatively flat continuum. The FWHM velocities ranged from few 100 \kms\ to 2000 \kms. The strong absorption lines or P-Cygni profiles which were seen in the optical spectra 10 days earlier were almost absent. The NIR spectrum was consistent with that of Fe II and hybrid class of classical novae, clearly indicating this event to be a classical nova.

ASASSN-18fv showed $\gamma$-ray emission \citep{jea18,pia18} with multiple peaks that were found to be coincident with optical flares \citep{Ayd20}. Hard X-rays were also detected during April 20-22 with the NuSTAR, while {\it Swift XRT} did not show any soft X-rays, \citep{nel18} indicating an internal shock. While no radio emission was reported at the nova position on April 3.3 \citep{ryd18}, synchrotron radio emission has been detected since $\sim 10$ days post the $\gamma$-ray detection \citep{Ayd20}.

In this paper, spectroscopic and morpho-kinematic structural evolution of the ejecta of ASASSN-18fv, based on the spectral data t = 8 to 316 days are presented. Here, t=0 corresponds to 2018 Mar 16, the date of discovery. 

\section{Observations}
The details of the optical spectroscopic data obtained from different instruments on several epochs are given in Table \ref{log}.
\begin{table*}
	\caption{Observational log for spectroscopic data obtained for ASASSN-18fv.}
	\label{log}
	\resizebox{0.8\hsize}{!}{\begin{tabular}{cccccc}
			\hline
			\hline
			& \textbf{Time since}  & \textbf{Exposure}  &  & \textbf{Wavelength}  & \\
			\textbf{Date} & \textbf{discovery} & \textbf{time} & \textbf{Resolution} & \textbf{range} & \textbf{Instrument}\\
			& \textbf{(days)} & \textbf{(s)} & & \textbf{(\AA)} & \\
			\hline
			2018 Mar 22 & 6 & 1800 & 15000 & 6500--6670 & LHIRES \\[0.25ex]
			2018 Mar 24 & 8 & 1800 & 15000 & 6500--6670 & LHIRES \\[0.25ex]
			2018 Mar 25 & 9 & 1800 & 15000 & 6500--6670 & LHIRES \\[0.25ex]
			2018 Mar 26 & 10 & 1800 & 15000 & 6500--6670 & LHIRES \\[0.25ex]
			2018 Mar 28 & 12 & 1200 & 15000 & 6500--6670 & LHIRES \\[0.25ex]
			2018 Mar 29 & 13 & 1800 & 15000 & 6500--6670 & LHIRES \\[0.25ex]
			2018 Mar 30 & 14 & 600 & 1500 & 3800--7250 & LISA \\[0.25ex]
			2018 Mar 30 & 14 & 1200 & 15000 & 6500--6670 & LHIRES \\[0.25ex]
			2018 Mar 31 & 15 & 600 & 1500 & 3800--7250 & LISA \\[0.25ex]
			2018 Apr 01 & 16 & 750 & 1500 & 3800--7250 & LISA \\[0.25ex]
			2018 Apr 01 & 16 & 1800 & 15000 & 6500--6670 & LHIRES \\[0.25ex]
			2018 Apr 03 & 18 & 1800 & 15000 & 6500--6670 & LHIRES \\[0.25ex]
			2018 Apr 04 & 18 & 2400 & 72000 & 4000-10000 & VBT Echelle \\[0.25ex]
			2018 Apr 04 & 19 & 420 & 1400 & 4250--7600 & OMR \\[0.25ex]
			2018 Apr 05 & 20 & 480 & 1300 & 4250--7600 & UAGS \\[0.25ex]
			2018 Apr 06 & 21 & 2400 & 27000 & 4000-10000 & VBT Echelle \\[0.25ex]
			2018 Apr 06 & 21 & 1200 & 1300 & 4250--7600 & UAGS \\[0.25ex]
			2018 Apr 07 & 22 & 1800 & 1300 & 4250--7600 & UAGS \\[0.25ex]
			2018 Apr 08 & 23 & 600 & 1500 & 3800--7250 & LISA \\[0.25ex]
			2018 Apr 09 & 24 & 1800 & 15000 & 6500--6670 & LHIRES \\[0.25ex]
			2018 Apr 09 & 24 & 300 & 1300 & 4250--7600 & UAGS\\[0.25ex]
			2018 Apr 10 & 25 & 180 & 1300 & 4250--7600 & UAGS\\[0.25ex]
			2018 Apr 11 & 26 & 1200 & 15000 & 6500--6670 & LHIRES \\[0.25ex]
			2018 Apr 30 & 45 & 1200 & 15000 & 6500--6670 & LHIRES \\[0.25ex]
			2018 May 05 & 50 & 500 & 1500 & 3800--8200 & LISA \\[0.25ex]
			2018 May 06 & 51 & 310 & 1500 & 3800--8200 & LISA \\[0.25ex]
			2018 May 09 & 54 & 1200 & 15000 & 6500--6670 & LHIRES \\[0.25ex]
			2018 May 10 & 55 & 210 & 1500 & 3800--8400 & LISA \\[0.25ex]
			2018 May 19 & 64 & 1200 & 15000 & 6500--6670 & LHIRES \\[0.25ex]
			2018 May 27 & 72 & 360 & 1500 & 3800--7400 & LISA \\[0.25ex]
			2018 Jun 21 & 96 & 360 & 1500 & 3800--7400 & LISA \\[0.25ex]
			2019 Jan 26 & 316 & 1224 & 66700 & 3800--8900 & SALT \\
			\hline
	\end{tabular}}
\end{table*}
\subsection{Mirranook Observatory}
Low and medium resolution optical spectroscopic observations were obtained from Mirranook observatory, Australia using two different spectrographs through a 0.28 m Celestron C11 telescope. The observatory is at 1100 m altitude and situated at Latitude 30$^\circ$ S giving good conditions to observe the target in Carina.
\subsubsection*{LISA}
LISA is a classical spectrograph with resolution R = 1500 covering a wavelength range of 3800--8000 \AA. Standard methods were used to reduce the data using various tasks in the Integrated Spectrographic Innovative Software (ISIS\footnote{\href{http://www.astrosurf.com/buil/isis-software.html}{http://www.astrosurf.com/buil/isis-software.html}}). Bias subtraction and extraction using the optimal extraction method were carried out on all the spectra. Wavelength calibration was done using the NeAr arc lamp spectrum. \textit{Medium-resolution Isaac Newton Telescope library of empirical spectra} (MILES) standard star \citep{san06} observed at similar airmass was used for instrument response correction. Simultaneous photometry of the target through $UBV$ filters using  American Association of Variable Star Observers (AAVSO\footnote{Kafka, S., 2018, Observations from the AAVSO International Database, \href{https://www.aavso.org}{https://www.aavso.org}}) comparison stars were carried out. These measurements were also submitted to the AAVSO database. 
\subsubsection*{LHIRES} 
Medium resolution spectra around H$\alpha$ were obtained using LHIRES Littrow spectrograph with a resolution of R = 15000. Data reduction was carried out using the standard methods in the ISIS. A neon calibration lamp taken before and after each image was used for calibration. Calibration was checked for accuracy using telluric lines visible near the H$\alpha$ wavelength.

\subsection{Vainu Bappu Observatory (VBO)}
\subsubsection*{UAGS and OMR}
Optical spectroscopic observations from the Universal Astro Grating Spectrograph (UAGS) mounted on the 1 m Carl Zeiss Telescope (CZT) and Opto Mechanics Research (OMR) spectrograph mounted on the 2.3 m Vainu Bappu Telescope (VBT) situated at VBO, Kavalur, India covering a range of 4200--7500 \AA\ with R = 1300 and 3800--8800 \AA\ with R = 1400 respectively were also obtained. Data reduction of these spectra like pre-processing, wavelength and flux calibration were carried out in the standard manner using Image Reduction and Analysis Facility (IRAF\footnote{IRAF is distributed by the National Optical Astronomy Observatory, which is operated by the Association of Universities for Research in Astronomy (AURA) under a cooperative agreement with the National Science Foundation.}). Wavelength calibration was carried out using FeAr and FeNe lamp spectra. Instrumental response correction was done using the spectrophotometric standard stars obtained on the same nights as that of the object. In order to convert the spectra to an absolute flux scale, zero points were obtained using $UBVRI$ magnitudes reported by AAVSO.

\subsubsection*{Echelle}
Optical spectroscopic observations were also carried out with the fiber-fed Echelle spectrograph at VBT covering range of 4000--10000 \AA\ with resolution R = 27000 and 72000. Standard procedures were followed to reduce the data using IRAF. Processes such as bad pixel removal, scattered light subtraction, bias corrections, flat-fielding and aperture extraction were executed. Wavelength calibration was done using ThAr lamp.

\subsection{South African Astronomical Observatory (SAAO)}
\subsubsection*{HRS}
The South African Large Telescope (SALT) High Resolution Spectrograph (HRS) \citep{bar08, bram10,bram12,cra14} was used to obtain spectral data on 2019 Jan 26. The dual beam fibre-fed echelle spectrograph, HRS covering the wavelength range of 3800--5500 \AA\ and 5450--8900 \AA in the high resolution (HR) mode at a resolution of R $\sim$ 66000 was used for observation. Pre-processing of the data like over-scan correction, bias subtraction and gain correction was conducted using the science pipeline of SALT \citep{craw10}. The remaining tasks like aperture extraction and wavelength calibration was carried out using the \textsc{echelle} package in IRAF.

\section{Analysis}
\subsection{Optical and NIR light curve, reddening and distance}
\begin{figure}
 \includegraphics[width=1\columnwidth]{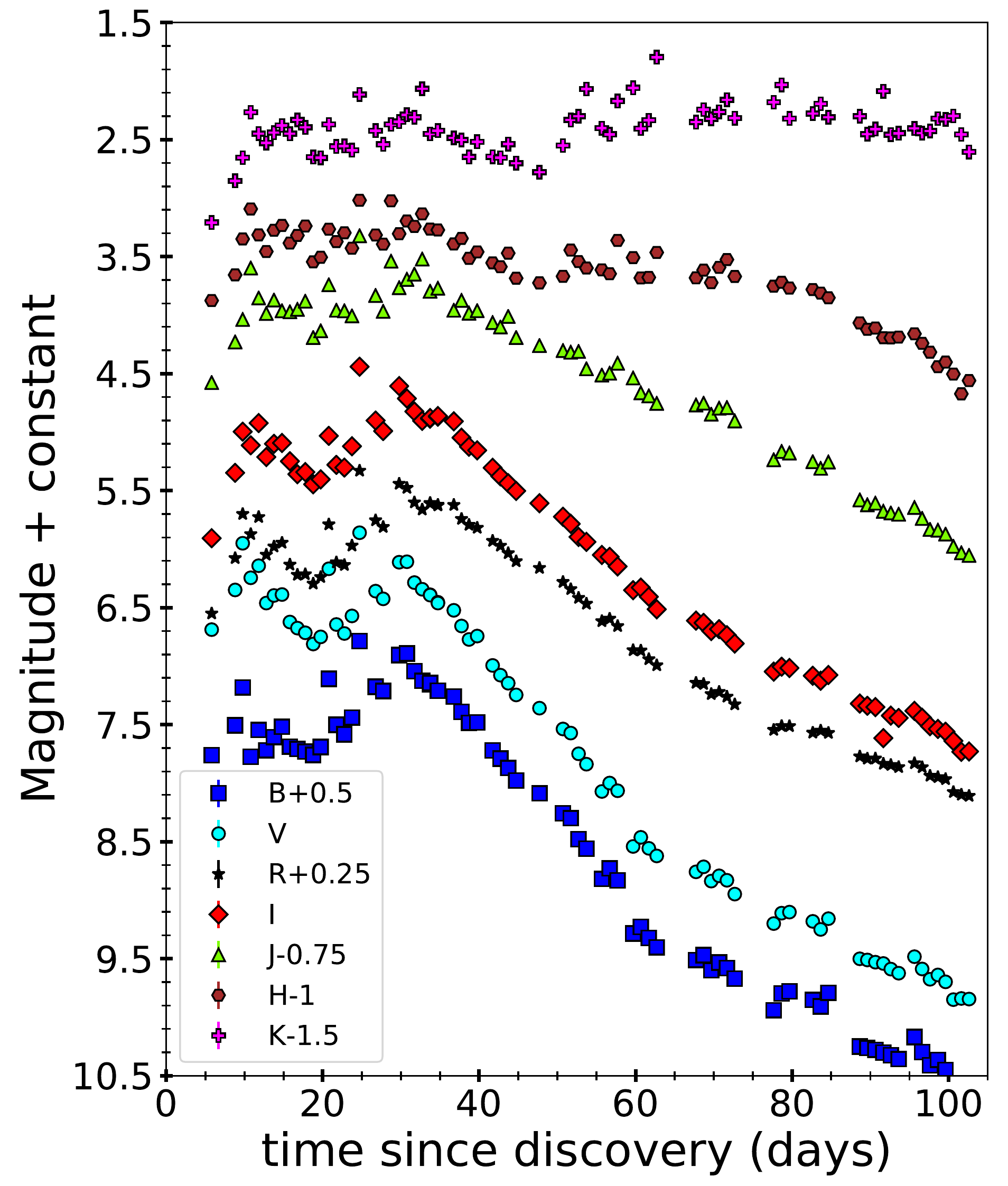}
 \caption{Apparent magnitude light curves of ASASSN-18fv generated using optical and NIR data from AAVSO and SMARTS. Offset have been applied for all the magnitudes except $V$ for clarity.}
 \label{lc}
\end{figure}

The optical and near-IR (NIR) light curves based on the data from AAVSO and Small and Moderate Aperture Research Telescope System (SMARTS\footnote{\href{http://www.astro.sunysb.edu/fwalter/SMARTS/NovaAtlas/}{http://www.astro.sunysb.edu/fwalter/SMARTS/NovaAtlas/}}) \citep{wal12} are presented in Fig. \ref{lc}.
The nova rose to maximum in about 8 days after discovery, reaching a peak brightness of $V\rm_{max}$ = 5.8. There was re-brightenings seen in the optical light curve from day 24--34 followed by a smoother decline. The NIR light curve followed a similar trend as optical with small fluctuations primarily in the H and K bands. From the least-squares regression fit to the post-maximum light curve, t$_2$ was estimated to be 50 $\pm$ 5 days and t$_3$ to be 70 $\pm$ 5 days. However, the t$_3$ value obtained using the relation t$_3$ = 2.75 (t$_2$)$^{0.88}$ by \cite{war95} is determined to be 85 days. This suggests that the system belongs to moderately fast novae category. The source was not detected on Mar. 15.34 with limiting magnitude of about 17.0 by \cite{sta18}, hence the amplitude of the nova was estimated as $\triangle$V $\sim$ 11 mag. In the amplitude versus decline rate plot for classical novae (Fig. 2.3 of \cite{war08}), t$_2$ for this system and the amplitude places the system closer to the center of observed spread. Using the maximum magnitude versus rate of decline (MMRD) relation by \cite{dow00}, the absolute magnitude of the nova was determined to be M$_V$ = -7.0 $\pm$ 0.1. Using this absolute magnitude value in the relation by \cite{liv92},  mass of white dwarf (WD) was estimated to be 0.74 M$_\odot$. Reddening was calculated using the observed spectrum and optical photometry data from SMARTS and AAVSO. \textit{E(B-V)} is found to be 0.76 $\pm$ 0.02 using the Na I line present in the observed spectrum, and determined using the relation given by \citet{Mun97}. \textit{E(B-V)} = 0.75 $\pm$ 0.07 and A$_V$ = 2.3 $\pm$ 0.2 for R = 3.1 are estimated using the intrinsic colors of novae at peak brightness \citep{van87}. The extinction map by \cite{nec80} shows A$_V$ value in the range of 1--2 in the direction of ASASSN-18fv, around 1--2 kpc. In a recent dust map given by \cite{sch11}, it was reported that the extinction value is A$_V$ $\sim$ 3.6 towards the direction of ASASSN-18fv, and a large error was suggested close to the Galactic center (l, b: 286, -1.1). The moderate value of A$_V$ estimated here appears to be reasonable even though the nova is located close to the direction of the Galactic centre. Using the above values, the distance to the nova was estimated to be d = 1.3 $\pm$ 0.2 kpc.

\begin{figure}
 \includegraphics[width=1\columnwidth]{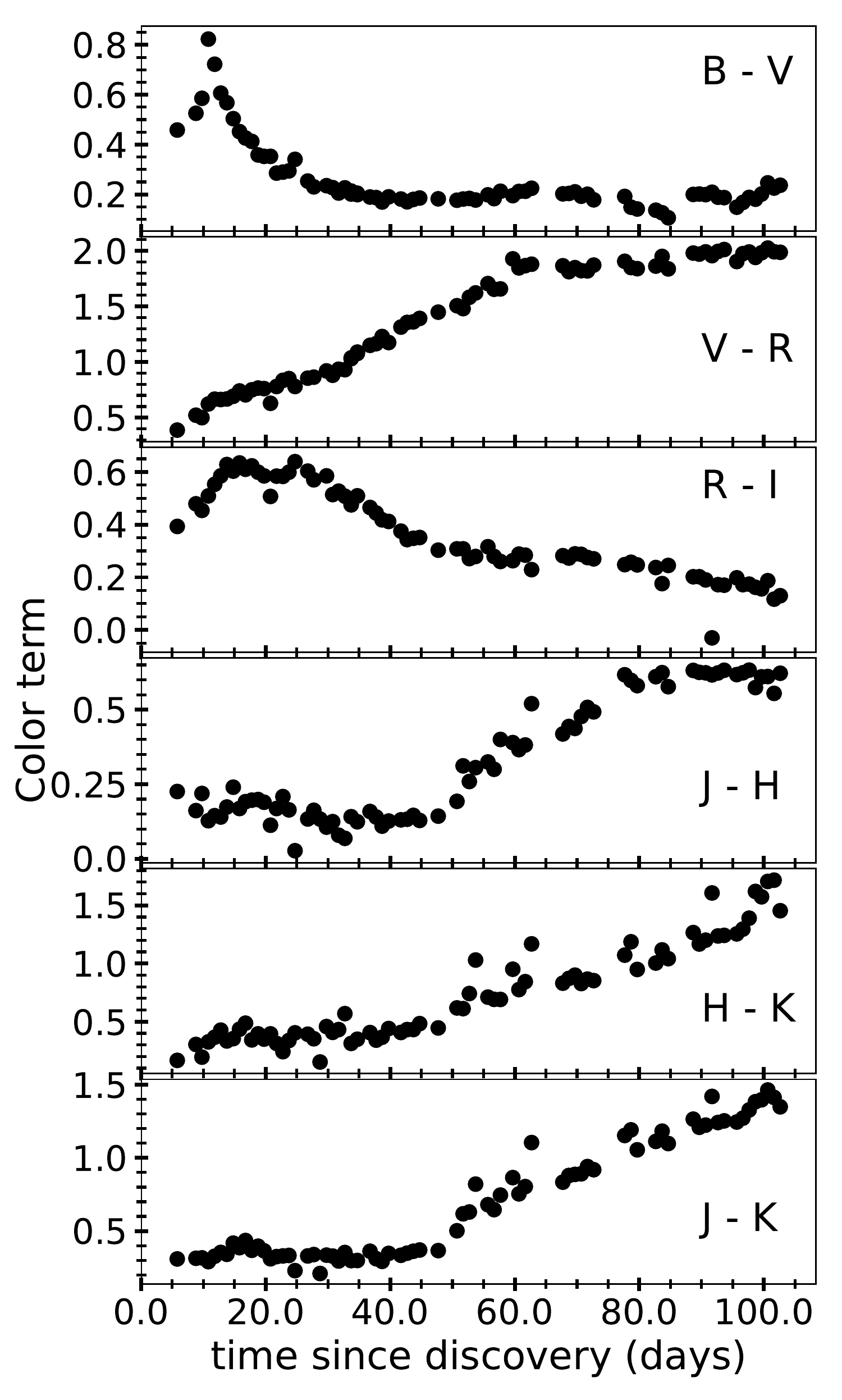}
 \caption{Evolution of optical and NIR color terms of nova ASASSN-18fv from day 0 (pre-maximum phase) to 100 (decline phase) since discovery.}
 	\label{colorterm}
\end{figure}

Optical and NIR color evolution for the nova is as shown in Fig. \ref{colorterm}. The $B$-$V$ and $R$-$I$ color evolution showed rise for first 10 days and then reached a peak value of 0.8 and 0.64 respectively. This was followed by a steep decline and flattening for $B$-$V$ and a slow decline for $R$-$I$ color. $J$-$H$, $H$-$K$, $J$-$K$ and $V$-$R$ color evolution showed increasing trend through out the observations with small fluctuations. The shape of the optical light curves of this nova (Fig. \ref{lc}) has the characteristics similar to that of a J class nova as it shows small fluctuations during its evolution. It can be classified as J(70) type as the estimated value of t$_3$ is 70 days (see \cite{str10} for more details).
\subsection{Spectral evolution}
\begin{figure*}
\begin{center}
\includegraphics[width=2\columnwidth]{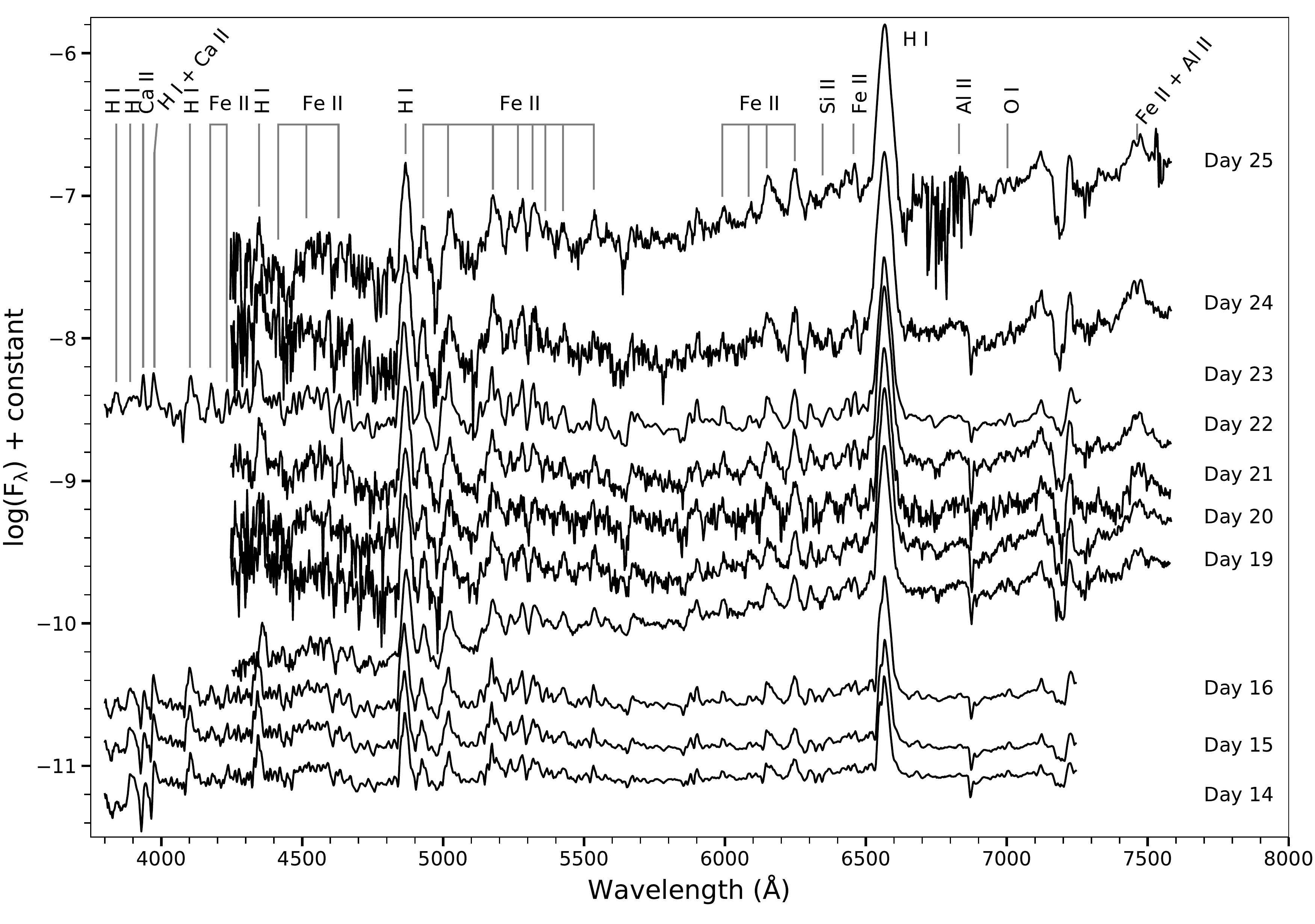}
	\caption{Low-resolution optical spectral evolution of ASASSN-18fv obtained from day 14 (2018 March 30) to day 25 (2018 April 10). Spectra are dominated by Fe II multiplets and hydrogen Balmer lines. The lines identified are marked, and time since discovery (in days) is marked against each spectrum.}
	\label{spectra1}
\end{center}
\end{figure*}

The low resolution spectral evolution of ASASSN-18fv during the pre-maximum and early decline phase is shown in Fig. \ref{spectra1}. The spectra show presence of hydrogen Balmer lines and Fe II multiplets along with Ca II (H and K) and He I lines at 4922, 5876 \AA. \cite{rab18} reported the NIR spectrum taken on day 16 (2018 April 1) and suggested that the spectrum is consistent with the normal Fe II and transition (Fe II + He/N) novae in the post maximum phase.  
\begin{figure*}
\begin{center}
\includegraphics[width=2\columnwidth]{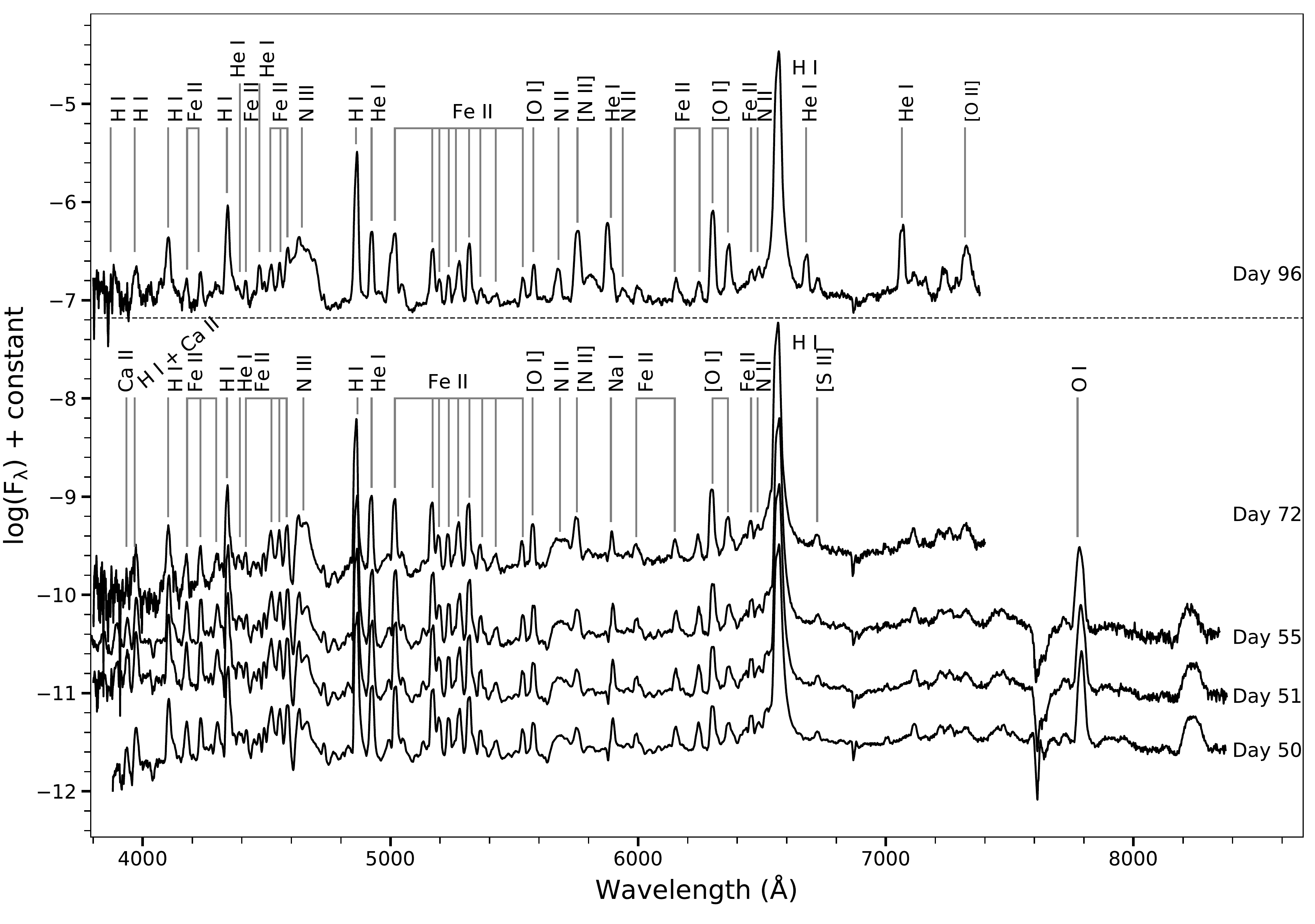}
	\caption{Low-resolution optical spectral evolution of ASASSN-18fv obtained from day 50 (2018 May 05) to day 96 (2018 June 21). From day 50 (2018 May 05) to 72 (2018 May 27), He, [O I], [N II] and N III lines appear and on day 96 (2018 June 21), He I lines become very prominent. The lines identified are marked, and time since discovery (in days) is marked against each spectrum.}
	\label{spectra2}
\end{center}
\end{figure*}

The spectrum of day 50 (2018 May 5) during the decline phase (Fig. \ref{spectra2}) consists of several emission lines like Fe II lines at 4179, 4297, 4549, 4629, 5018, 5169, 5284 \AA\ etc. along with [N II] 5755, [O I] 5577, 6300 and 6364, [S II] 6724, O I 7002, 7477. The other emission lines present were He I 4471, 5048 \AA\ and broad He II 8237 \AA\ and C II 4267 \AA\ and N II 6482 \AA. The spectrum obtained on day 96 (2018 June 21) consists of few Fe II lines along with some higher excitation lines like He I 4922, 5876, 6678, 7065 and 7281 \AA, Fe II lines at 5007 and 5018 \AA, C IV 5805 \AA, N II 5938, 6482 \AA. Hybrid nature of this system was suggested by \cite{rab18} from the NIR spectrum (0.8--1.7 $\mu$) and the lines like H I, He I, Ca II, N I, O I, C I were identified.  The optical spectra of ASASSN-18fv also showed the presence of C II 4267 \AA\ and N II 6482 \AA\ lines and flat top profiles in 2018 May and June. The presence of spectral lines of He, N and C II at such early epochs and also the flat top profiles \citep{wil92} suggest that ASASSN-18fv belongs to the hybrid class of nova.

High resolution spectrum was obtained on day 316 (2019 Jan 26) using SALT HRS spectrograph. The prominent lines present in the spectrum are hydrogen Balmer, N III 4638 \AA\, He II 4686 \AA, He I 5876, 6678 and 7068 \AA and [N II] 5755 \AA. The emission line profiles are structured with multiple peaks and are boxy. The He I lines are boxy with multiple peaks. The H$\beta$ line profile appears to have boxy profile with three peaks.

\begin{figure}
	\includegraphics[width=\columnwidth]{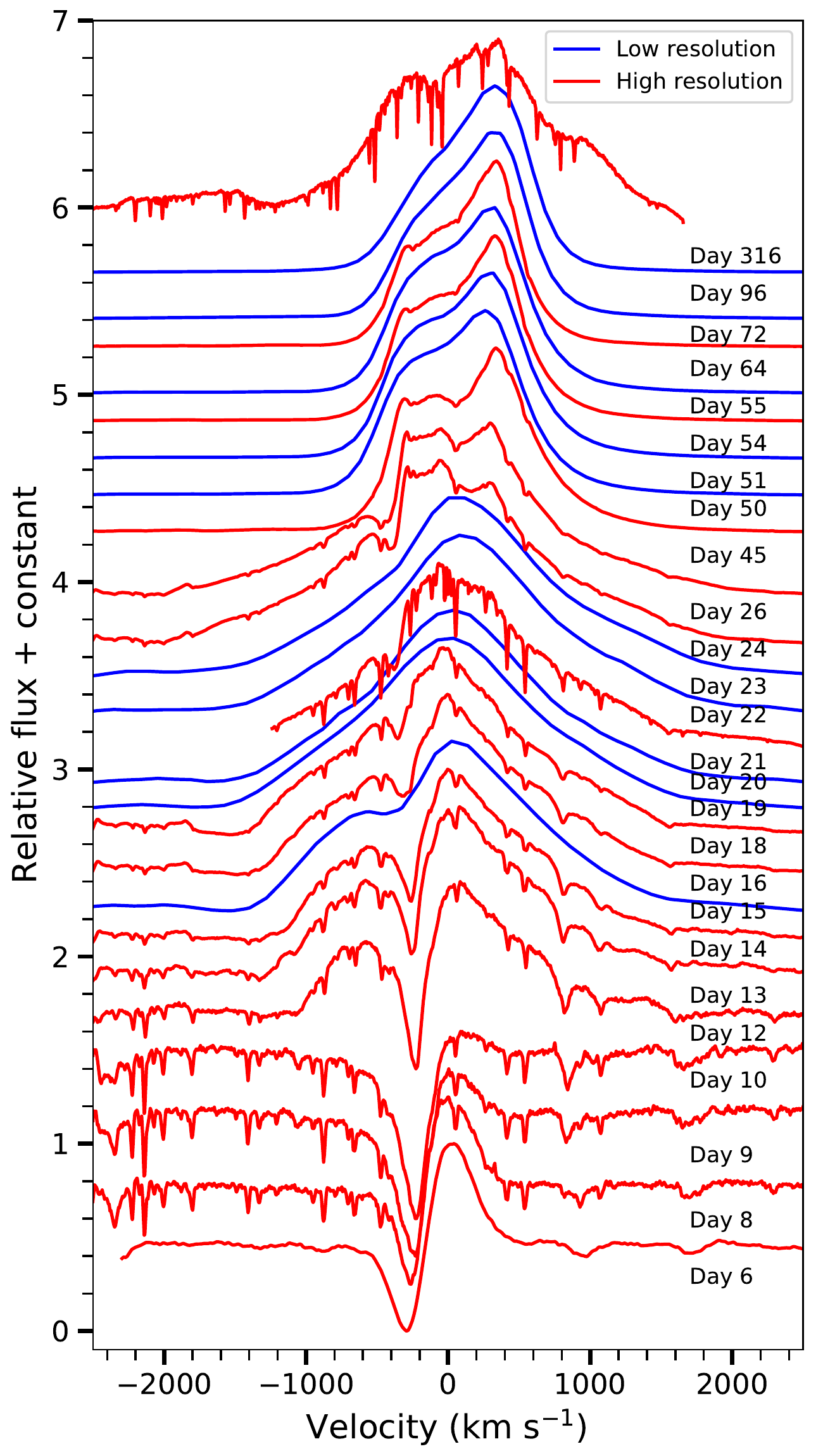}
	\caption{Evolution of H$\alpha$ velocity profile of ASASSN-18fv from day 6 (2018 Mar 22) to 316 (2019 Jan 26) obtained using the low, medium and high resolution spectroscopic data. The line profiles evolve from P-Cygni in the pre-maximum phase to a boxy and structured one in the decline phase.}
	\label{halpha}
\end{figure}
\begin{figure}
	\includegraphics[width=1\columnwidth]{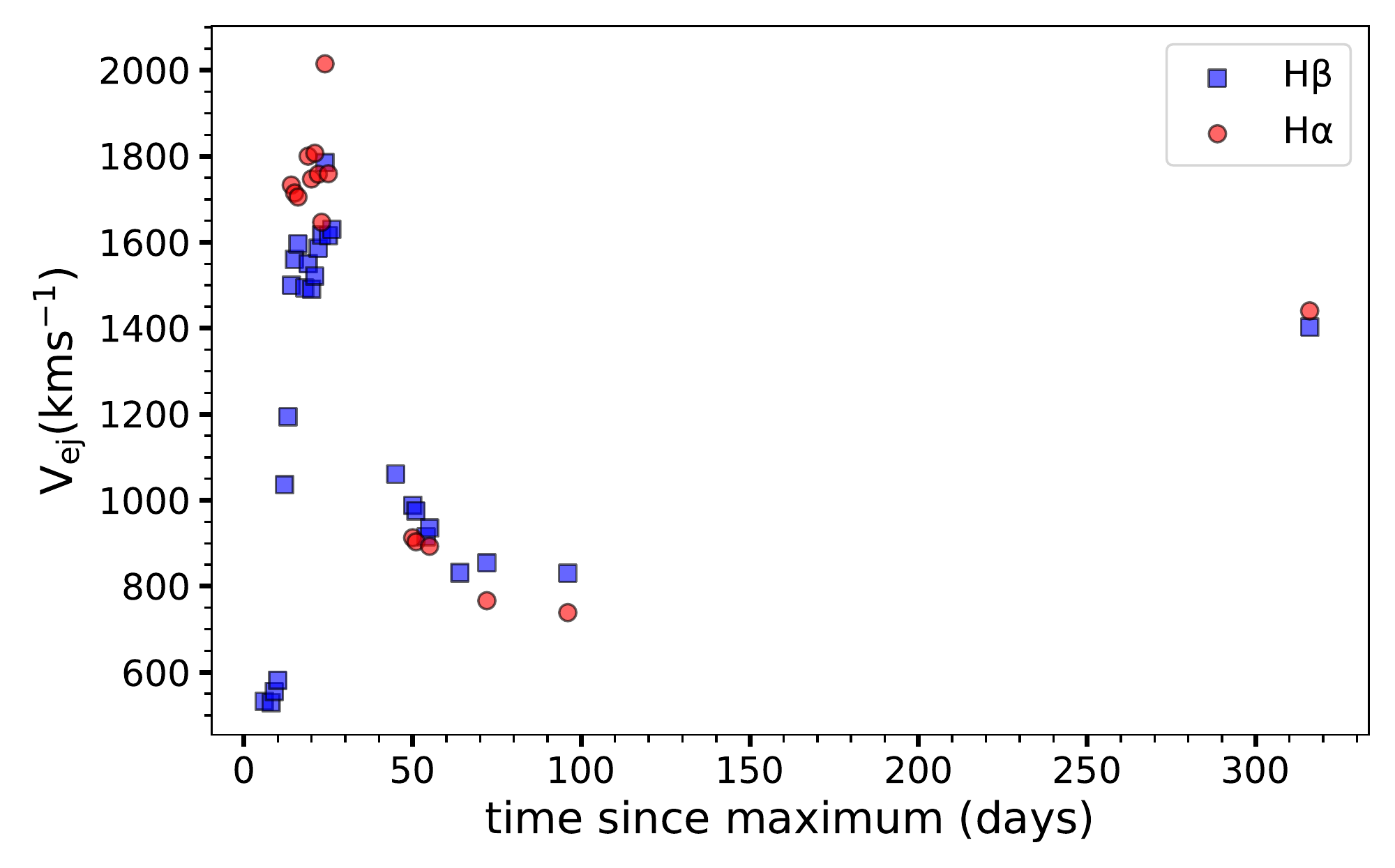}
	\caption{Evolution of H$\alpha$ and H$\beta$ FWHM velocities of ASASSN-18fv from day 14 (2018 Mar 30) to day 96 (2018 Jun 21).}
	\label{fwhm}
\end{figure}

The H$\alpha$ velocity profile evolution is as shown in Fig. \ref{halpha}. The P-Cygni profile of H$\alpha$ line has blue-shifted component at $\sim$ -250 \kms. The P-Cygni profile was present till day 14. The observed FWHM velocities of H$\alpha$ and H$\beta$ lines are plotted in Fig. \ref{fwhm}. During the early decline phase, the velocities range from 1500--2000 \kms. In the decline phase, the FWHM velocities range from 800--1100 \kms\ followed by an increase in the velocities to 1400 \kms\ on day 316.

\subsection{Physical parameters}
Using the line fluxes of hydrogen and oxygen from the optical spectrum physical parameters like the hydrogen mass, optical depth of oxygen and electron temperature can be estimated. Further, modelling the spectrum using photo-ionization analyses can provide estimates of the  source luminosity, effective temperatures, density and chemical abundances. We use here the spectrum of day 96 (2018 June 21) to obtain the various physical parameters of the nova material. Day 96 is chosen as this spectrum is during the transition phase, showing both Fe II and the He/N features.

The oxygen optical depth and electron temperature are estimated using the formulae by \cite{wil94}, 
\begin{align}
\label{op_dep}
\dfrac{F_{\lambda6300}}{F_{\lambda6364}} = \dfrac{(1 - e^{-\tau})}{(1-e^{-\tau/3})}.
\end{align}
Using the value of $\tau$, the electron temperature is given  by
\begin{align}
\label{Te}
T_e =   \dfrac{11200}{ log [\dfrac{(43\tau)}{(1-e^{-\tau})} \times \dfrac{F_{\lambda6300}}{F_{\lambda5577}}]}
\end{align}
where F$_{\lambda5577}$, F$_{\lambda6300}$ and F$_{\lambda6364}$ are the line intensities of [O I] 5577, 6300 and 6364 \AA\ lines. The optical depth $\tau$ of the ejecta for [O I] 6300 \AA\ is 0.50 $\pm$ 0.13. The electron temperature T$_e$ calculated using the estimated value of $\tau$ is found to be 7026 $\pm$ 74 K.

The hydrogen mass can also be estimated using the relation by \cite{ost06},
\begin{align}
	\dfrac{m(H)}{M_{\odot}} = d^{2} \times 2.455 \times 10^{-2} \times \dfrac{I(H\beta)}{\alpha_{eff}N_e}
\end{align}
 where $\alpha_{eff}$ is the effective recombination coefficient obtained from \cite{hum95} and I(H$\beta$) is the flux of H$\beta$ line. The hydrogen mass thus determined is found to be (2.16 $\pm$ 0.17) $\times$ 10$^{-4}$ M$_{\odot}$.
 
The photo-ionization code CLOUDY, C17.01 \citep{Fer17} is used to model this system to understand the physical conditions of the system like source luminosity, effective temperature, density and elemental abundances. A detailed description for obtaining the 1D photo-ionization model is available at \cite{Raj18} and \cite{Pav19}.

\begin{table}
	\caption{Observed and best-fit CLOUDY model line flux ratios for day 96 of ASASSN-18fv}
	\label{d96_c}
	\begin{center}
		\resizebox{\hsize}{!}{%
			\begin{tabular}{lclcc}
				\hline
				\hline
				\textbf{Line ID} & \boldmath{$\lambda$ (\AA)} & \textbf{Observed}$^{a}$ & \textbf{Modelled}$^{a}$ & \boldmath{$\chi^2$}  \\
				\hline
				H I           & 3889       & 1.57E$-$01 & 5.02E$-$01 & 2.94E$+$00 \\
				H I           & 3970       & 2.31E$-$01 & 5.09E$-$01 & 1.03E$+$00 \\
				He I          & 4026       & 4.92E$-$02 & 3.52E$-$01 & 9.23E$-$01 \\
				H I           & 4102       & 6.89E$-$01 & 8.94E$-$01 & 1.39E$+$00 \\
				Fe II         & 4179       & 7.55E$-$02 & 2.21E$-$01 & 3.01E$+$00 \\
				Fe II         & 4233       & 1.21E$-$01 & 4.47E$-$01 & 1.04E$+$00 \\
				H I           & 4340       & 1.01E$+$00 & 1.34E$+$00 & 1.21E$+$00 \\
				Fe II         & 4352       & 2.28E$-$01 & 6.76E$-$01 & 2.56E$+$00 \\
				He I          & 4471       & 2.77E$-$01 & 3.75E$-$01 & 4.00E$-$01 \\
				Fe II + N III & 4517       & 4.85E$-$01 & 8.29E$-$01 & 1.94E$+$00 \\
				Fe II         & 4556       & 1.41E$+$00 & 1.54E$+$00 & 3.14E$+$00 \\
				Fe II         & 4584       & 1.03E$+$00 & 9.37E$-$01 & 5.60E$-$01 \\
				N III         & 4638       & 4.01E$+$00 & 4.06E$+$00 & 3.87E$-$02 \\
				He II         & 4686       & 7.02E$-$01 & 8.39E$-$01 & 6.94E$-$01 \\
				H I           & 4861       & 1.00E$+$00 & 1.00E$+$00 & 0.00E$+$00 \\
				He I + Fe II  & 4924       & 4.28E$-$01 & 6.49E$-$01 & 1.10E$+$00 \\
				He I + Fe II  & 5016       & 1.11E$-$01 & 9.91E$-$02 & 1.53E$-$03 \\
				He I          & 5048       & 2.82E$-$01 & 5.59E$-$01 & 1.08E$+$00 \\
				Fe II         & 5169       & 4.65E$-$01 & 6.64E$-$01 & 1.47E$+$00 \\
				Fe II         & 5198       & 4.43E$-$02 & 1.93E$-$01 & 4.92E$-$01 \\
				Fe II         & 5235       & 4.53E$-$05 & 1.64E$-$01 & 8.50E$-$01 \\
				Fe II         & 5276       & 1.15E$-$01 & 1.91E$-$01 & 8.64E$-$02 \\
				Fe II         & 5317       & 2.22E$-$01 & 6.06E$-$01 & 1.72E$+$00 \\
				Fe II         & 5363       & 2.82E$-$01 & 1.46E$-$01 & 3.50E$-$01 \\
				He II         & 5412       & 1.09E$+$00 & 1.18E$+$00 & 9.59E$-$02 \\
				Fe II         & 5535       & 1.02E$+$00 & 1.37E$+$00 & 2.20E$+$00 \\
				{[}O I{]}         & 5577       & 9.72E$-$02 & 1.06E$-$01 & 3.61E$-$03 \\
				N II          & 5679       & 2.66E$-$03 & 2.87E$-$02 & 2.52E$-$02 \\
				{[}N II{]}        & 5755       & 1.39E$-$01 & 3.27E$-$02 & 5.40E$-$01 \\
				C IV          & 5805       & 9.45E$-$02 & 1.39E$-$01 & 5.26E$-$01 \\
				He I          & 5876       & 1.39E$+$00 & 1.94E$+$00 & 3.00E$+$00 \\
				{[}O I{]}         & 6300       & 1.43E$+$00 & 1.15E+00 & 1.01E$+$00 \\
				{[}O I{]}         & 6364       & 5.53E$-$01 & 2.39E$-$01 & 2.36E$+$00 \\
				H I           & 6563       & 3.06E$-$01 & 3.91E$-$01 & 1.67E$+$00 \\
				He I          & 6678       & 1.11E$+$00 & 1.20E+00 & 1.00E$-$01 \\
				He I          & 7065       & 2.03E$-$01 & 1.09E$-$01 & 1.11E$-$01 \\
				C II          & 7235       & 3.57E$-$01 & 4.34E$-$01 & 4.00E$-$01 \\
				{[}O II{]}  & 7320-30    & 1.00E$+$00 & 6.64E$-$01 & 1.85E$+$00 \\
				\hline
		    \end{tabular}}
	\end{center}
	$^{a}$Relative to H$\beta$
\end{table}
\begin{table}
	\caption{Best-fit CLOUDY model parameters obtained on day 96 for the system ASASSN-18fv}
	\label{d96_cp}
	\begin{center}
		\resizebox{\hsize}{!}{%
			\begin{tabular}{l c } 
				\hline\hline
				\textbf{Parameter} & \textbf{Day 96} \\ [0.5ex] 
				\hline 
				T$_{BB}$ ($\times$ 10$^{5}$ K) & 1.38 $\pm$ 0.10 \\ [0.25ex]
				Luminosity ($\times$ 10$^{38}$ erg/s) & 1.00 $\pm$ 0.08 \\ [0.25ex]
				Clump Hydrogen density ($\times$ 10$^{11}$ cm$^{-3}$) & 1.58  \\ [0.25ex]
				Diffuse Hydrogen density ($\times$ 10$^{9}$ cm$^{-3}$) & 3.16 \\ [0.25ex]
				Covering factor (clump) & 0.80 \\ [0.25ex]
				Covering factor (diffuse) & 0.20 \\ [0.25ex]
				$\alpha$ & -3.00 \\ [0.25ex]
				Inner radius ($\times$ 10$^{14}$ cm) & 1.45 \\ [0.25ex]
				Outer radius ($\times$ 10$^{14}$ cm) & 6.76 \\ [0.25ex]
				Filling factor  & 0.10 \\ [0.25ex]
				N/N$_{\odot}$ & 3.48 $\pm$ 0.14 (4)$^{a}$ \\ [0.25ex]
				O/O$_{\odot}$ & 2.88 $\pm$ 0.11 (4) \\ [0.25ex]
				Fe/Fe$_{\odot}$ & 2.20 $\pm$ 0.10 (15) \\ [0.25ex]
				He/He$_{\odot}$ & 2.14 $\pm$ 0.04 (9) \\ [0.25ex]
				C/C$_{\odot}$ & 1.00 $\pm$ 0.03 (2) \\ [0.25ex]
				Ejected mass ($\times$ 10$^{-4}$ M$_{\odot}$) & 6.07 \\ [0.25ex]
				Number of observed lines (n) & 38 \\ [0.25ex]
				Number of free parameters (n$_{p}$) & 12 \\ [0.25ex]
				Degrees of freedom ($\nu$) & 26 \\ [0.25ex]
				Total $\chi^{2}$ & 41.92 \\ [0.25ex]
				$\chi^{2}_{red}$ & 1.61 \\ [0.25ex]
				\hline
		\end{tabular}}
	\end{center}
	$^{a}$The number of lines availed to obtain abundance estimate is as shown in the parenthesis.
\end{table}

\begin{figure}
	\includegraphics[scale=0.35]{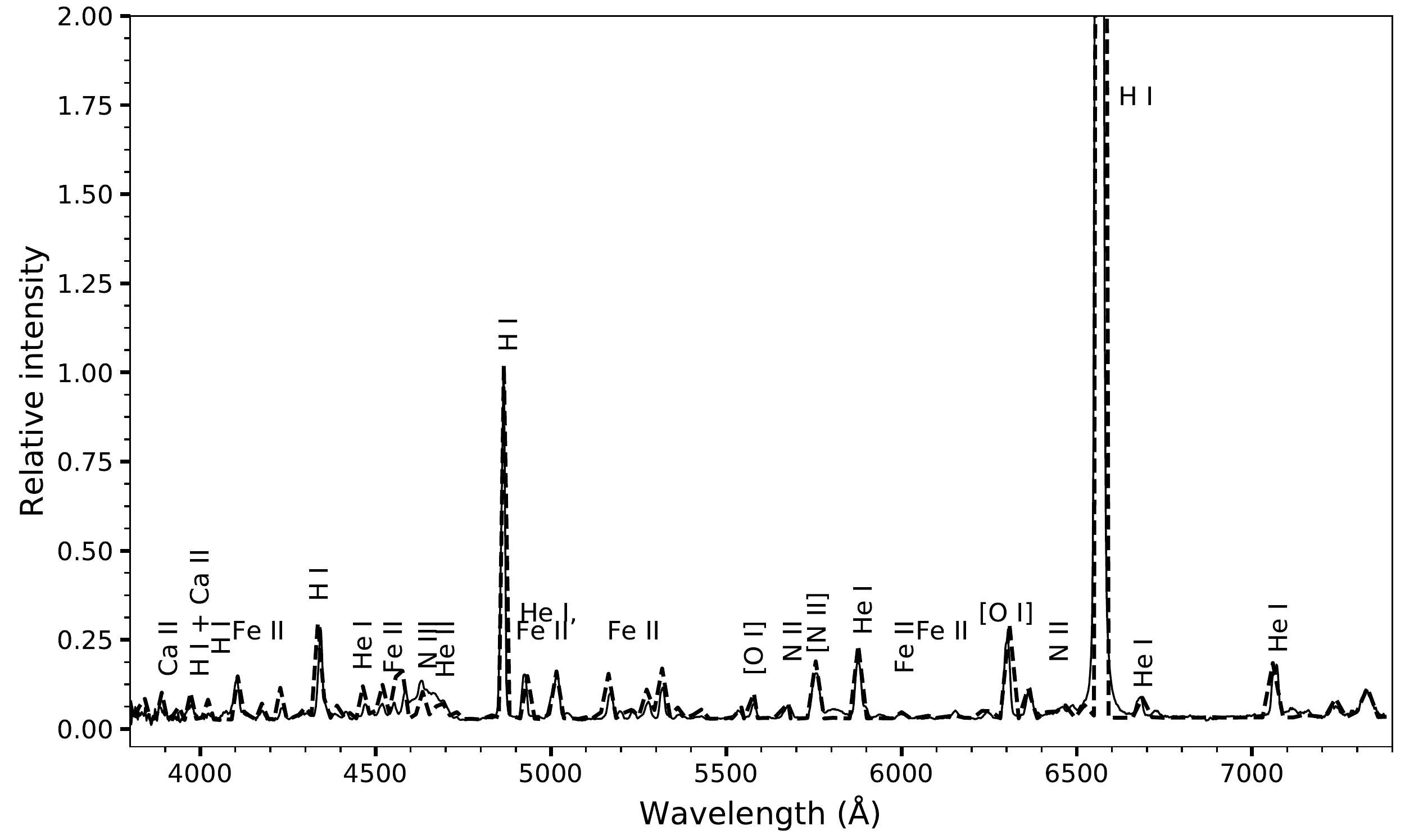}
	\caption{Best-fit CLOUDY synthetic spectrum (dash line) plotted over the observed spectrum (continuous line) of ASASSN-18fv obtained on 2018 June 21 (day 96).}
	\label{d96_1d}
\end{figure}

Three regions were used to model the observed spectrum. Fe II and He lines were generated by two different clump components, and [N II] (5755 \AA), [O I] (5577, 6300, 6364 \AA), [O II] (7320 \AA) and N II (5679 \AA) lines were generated by a diffuse component (low density) covering 15\% of the volume. Several spectra are produced in order to obtain the best-fit one by varying free parameters such as hydrogen density, effective blackbody temperature, and abundances of the elements based on the lines present in the observed spectrum. The best-fit synthetic spectrum was obtained by calculating the $\chi^2$ using the equation 2 in \cite{Pav19}. The effective temperature and luminosity for the central ionizing source are found to be 1.38 $\times$ 10$^{5}$ K and 10$^{38}$ erg s$^{-1}$ respectively. The clump hydrogen density is 10$^{10}$ cm$^{-3}$ and diffuse hydrogen density, 3.16 $\times$ 10$^{9}$ cm$^{-3}$. The relative fluxes of the observed and modelled lines, and their corresponding $\chi^{2}$ values are given in Table \ref{d96_c}. The best-fit parameter values obtained from the model are given in Table \ref{d96_cp}. The best-fit abundance values show that helium, iron, oxygen and nitrogen are over abundant as compared to the solar values, while carbon has a solar abundance value. The ejected mass was estimated using the equation 3 in \cite{Pav19} and found to be 6.07 $\times$ 10$^{-4}$ M$_{\odot}$. The best-fit 1D spectrum obtained is as shown in Fig. \ref{d96_1d}. The optical depth $\tau$ of the ejecta for [O I] 6300 \AA\ is 0.57 and the electron temperature T$_e$ is $\sim$6000 K from the photo-ionization 1D model and these values are consistent with that obtained from spectral analysis.

\subsection{Morpho-kinematic analysis} 
The morpho-kinematic application, SHAPE \citep{ste06} is used to analyze and understand the morpho-kinematic geometry of the nova ejecta using velocity profiles. Morpho-kinematic analysis of the velocity profiles of the emission features provide information about the inclination angle, position angle and geometrical structure of the ejecta. Similar work has been done earlier by \cite{Mun11} for V2672 Oph where the overall structure of the ejecta was found to be prolate system with polar blobs and an equatorial ring, and also by \cite{Rib11} for V2491 Cyg where the remnant morphology was found to be of polar blobs and an equatorial ring. As hydrogen is the most abundant element in novae, the model profiles corresponding to the observed profiles are obtained for epochs like day 18, 21, 64, 96 and 316. This also reveals the evolution of geometry of the ejecta. Few other line profiles like He I, He II, O I, N III, and [N II] are also modelled to study the behaviour of the ejecta geometry. 

\begin{figure*}
	\includegraphics[width=2\columnwidth]{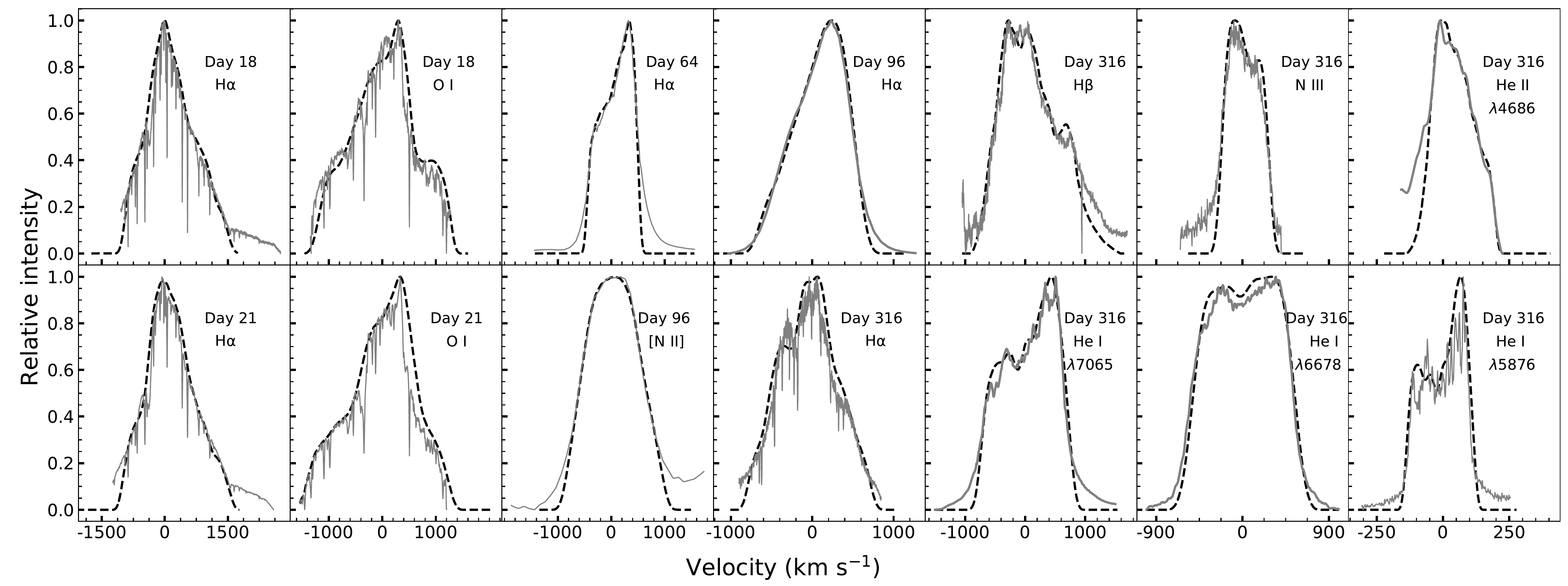}
	\caption{Best-fit modelled velocity profiles (dash black line) plotted over the observed H$\alpha$ profile (continuous grey line) of ASASSN-18fv obtained on days 18, 21, 64, 96 and 316 since discovery. The epochs and lines are marked against each profile.}
	\label{shp_prof}
\end{figure*}

Initially the ejecta is assumed to be in the form of spheroidal (or ellipsoidal) shell with (or without) equatorial rings. All the components are further defined with necessary modifiers such as velocity and density. For some components, the squeeze modifier is also defined when necessary. The radius of all the components are defined using the FWHM of the observed emission line and time since outburst. The density values are calculated using the observed fluxes of the emission lines and velocity using the FWHM of the emission line. These values are used to define the density and velocity modifiers. Initially, a synthetic line profile close to that of the observed one is obtained through eye by varying the components. The initial components of the nova shell were defined based on studies of nova shell models in the literature, such as \cite{Hut72} and \cite{Gil99}. In subsequent iterations, the values in the squeeze modifier, inclination and position angles are varied until the best-fit profile is obtained. The flux density of the modelled profile is compared with that of the observed profile. The best-fitting modelled profile is obtained by using the same relation for $\chi^2$ as that of the photo-ionization analysis. The best-fit model profiles over-plotted on the observed profiles are as shown in Fig. \ref{shp_prof}. The details of this analysis such as used and obtained parameters are as provided in Table \ref{shp_param}.

\begin{table}
\caption{Details of the morpho-kinematic analysis of the nova shell of ASASSN-18fv at different epochs.}
\label{shp_param}
\resizebox{\hsize}{!}{%
\begin{tabular}{cccccc}
\hline
\hline
\textbf{Epoch} & \textbf{Line profile} & \textbf{Size ratio$^a$} & \textbf{i$^b$ ($^\circ$)} & \textbf{P.A.$^c$ ($^\circ$)} & \boldmath{$\chi_{red}^2$$^d$} \\
\hline
\multirow{2}{*}{Day 18} & H$\alpha$ & 3.71 $\pm$ 0.12 & 55 $\pm$ 1.65 & 83 $\pm$ 1.85 & 1.39 \\
& O I 8446 \AA & 4.21 $\pm$ 0.10 & 60 $\pm$ 1.80 & 83 $\pm$ 1.70 & 1.76 \\ \hdashline
\multirow{2}{*}{Day 21} & H$\alpha$ & 3.79 $\pm$ 0.11 & 53 $\pm$ 1.55 & 83 $\pm$ 1.60 & 1.46 \\
 & O I 8446 \AA & 4.79 $\pm$ 0.12 & 60 $\pm$ 1.75 & 83 $\pm$ 1.45 & 1.82 \\ \hdashline
Day 64 & H$\alpha$ & 1.68 $\pm$ 0.04 & 53 $\pm$ 1.70 & 82 $\pm$ 1.85 & 1.24 \\ \hdashline
\multirow{2}{*}{Day 96} & [N II] 5755 \AA & 1.90 $\pm$ 0.02 & 50 $\pm$ 1.25 & 82 $\pm$ 1.70 & 1.22 \\
 & H$\alpha$ & 1.45 $\pm$ 0.03  & 53 $\pm$ 1.65 & 82 $\pm$ 1.55 & 1.11 \\ \hdashline
\multirow{7}{*}{Day 316} & N III 4638 \AA & 3.06 $\pm$ 0.07 & 50 $\pm$ 1.10 & 85 $\pm$ 1.35 & 1.73 \\
 & He II 4686 \AA & 3.13 $\pm$ 0.04 & 40 $\pm$ 1.30 & 85 $\pm$ 1.00 & 1.81 \\
 & H$\beta$ & 1.73 $\pm$ 0.06 & 50 $\pm$ 1.50 & 85 $\pm$ 1.15 & 1.62 \\
 & He I 5876 \AA & 2.65 $\pm$ 0.11 & 35 $\pm$ 1.65 & 85 $\pm$ 1.60 & 1.77 \\
 & H$\alpha$ & 3.18 $\pm$ 0.05 & 50 $\pm$ 1.95 & 85 $\pm$ 1.25 & 1.86 \\
 & He I 6678 \AA & 2.09 $\pm$ 0.04 & 40 $\pm$ 1.35 & 85 $\pm$ 1.00 & 1.41 \\
& He I 7065 \AA & 2.64 $\pm$ 0.04 & 35 $\pm$ 1.40 & 85 $\pm$ 1.80 & 1.58 \\
\hline
\end{tabular}}
$^a$ Ratio of polar to the equatorial size.\\
$^b$ Inclination angle of the ejecta axis. \\
$^c$ Position angle of the ejecta axis. \\
$^d$ Reduced $\chi^2$ calculated using the observed and modelled profiles and the degrees of freedom in the model.
\end{table}

\subsubsection*{Days 18 and 21}
\begin{figure*}
    \centering
    \includegraphics[width=2\columnwidth]{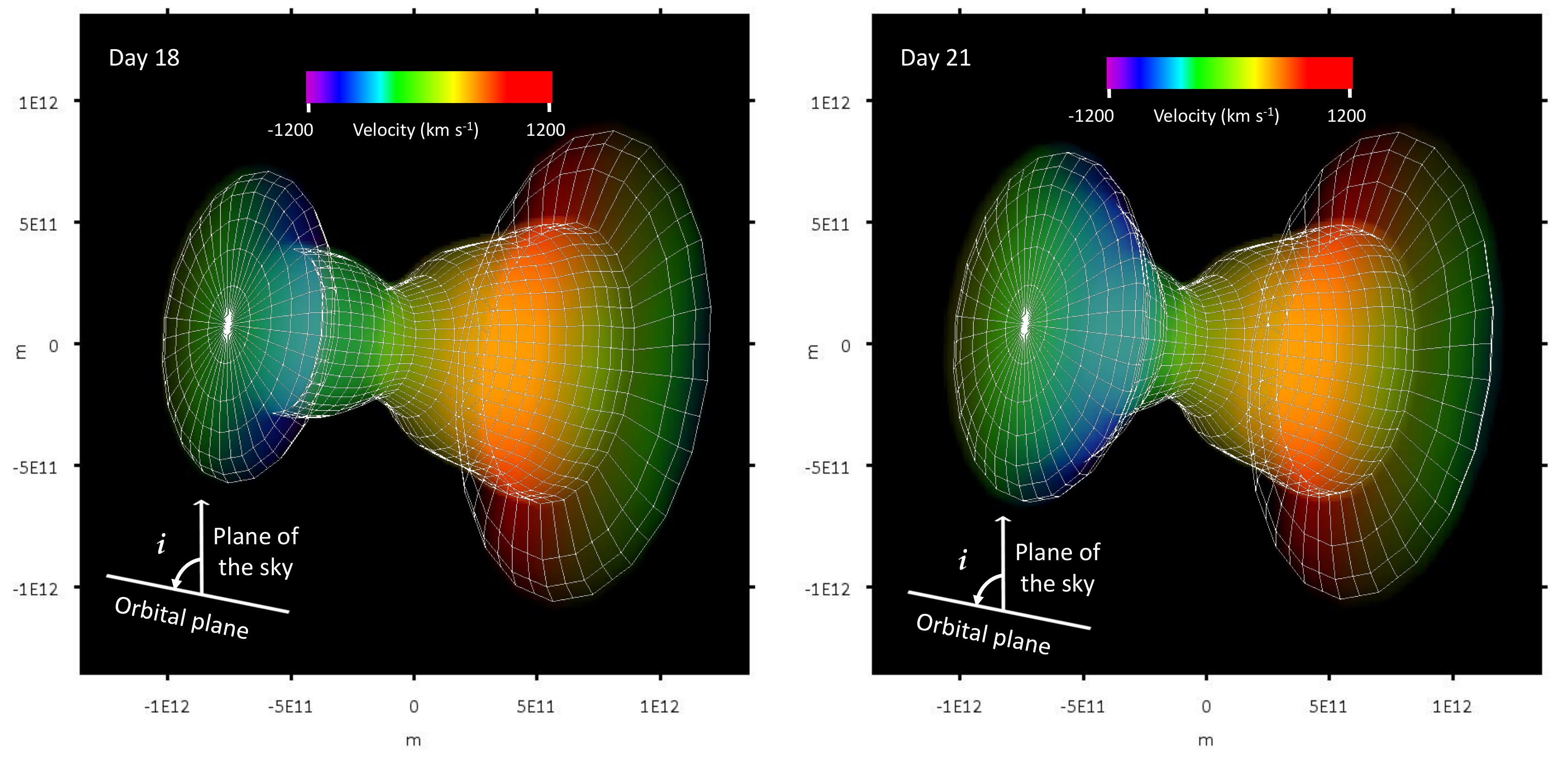}
    \caption{The asymmetric dumbbell-like O I structure of the ejecta of ASASSN-18fv in the two-dimensional plane with X-axis being the line-of-sight direction and Y being the axis perpendicular to that of the plane of sky and line-of-sight. The colour gradient represents the velocity values varying along the line-of-sight. The structures are obtained using the velocity profile of O I 8446 \AA\ obtained on 2018 April 03 (day 18) and 2018 April 06 (day 21). Here, the inclination angle of the system is the angle between the plane of the sky and orbital plane.}
    \label{o1_shape}
\end{figure*}

The ejecta geometry of O I 8446 \AA\ obtained on day 18 and 21 (2018 April 03 and 06) is an asymmetric bipolar one (dumbbell shape) with an inclination angle of about 60$^\circ$. The best-fit O I 8446 \AA\ geometrical structures corresponding to their best-fit velocity profiles are as shown in Fig. \ref{o1_shape}. 

\begin{figure*}
    \centering
    \includegraphics[width=2\columnwidth]{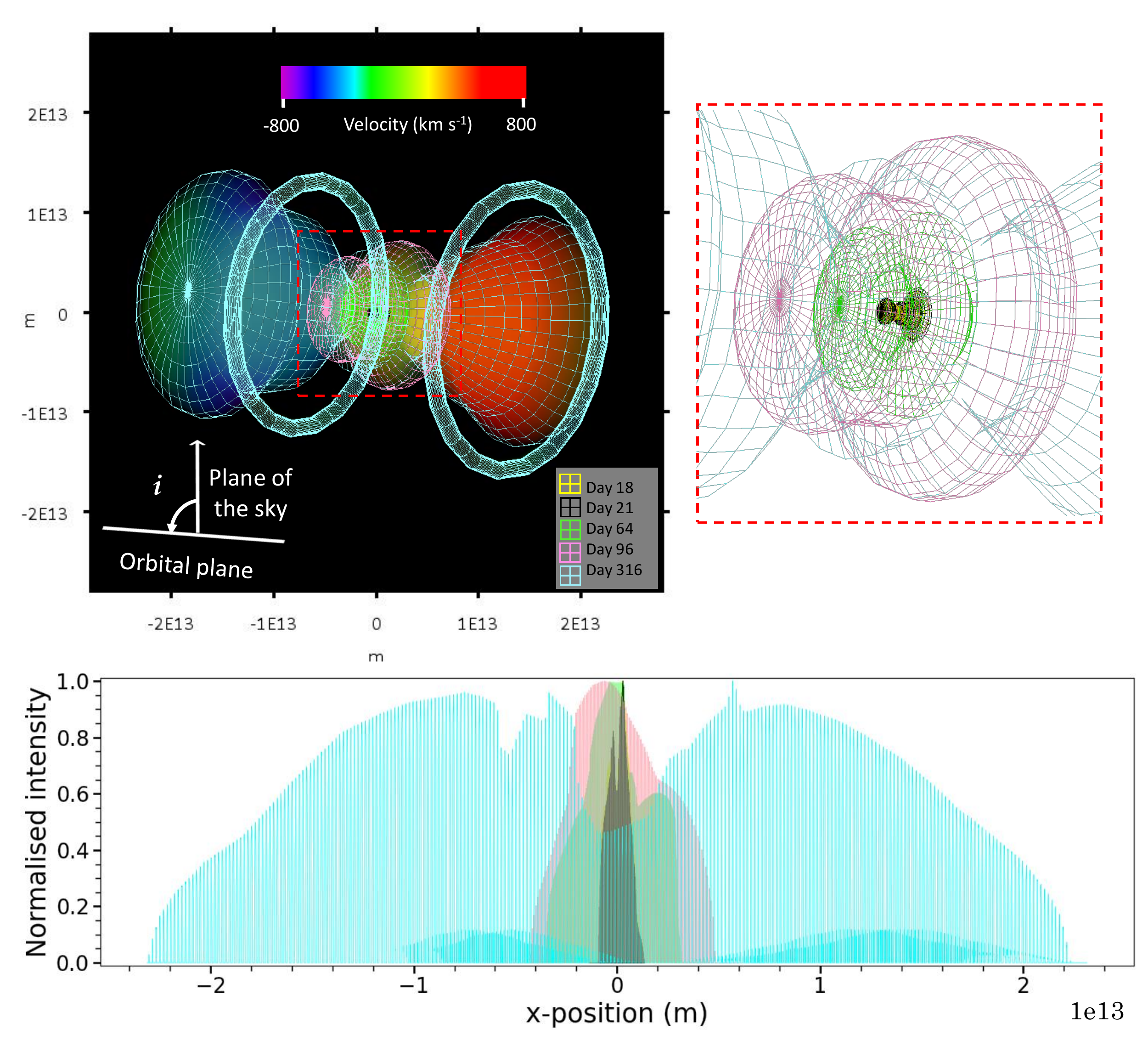}
    \caption{\textit{Top-left:} The dumbbell-like H$\alpha$ geometry obtained on days 18, 21, 64, 96 and 316 using the H$\alpha$ velocity profiles of ASASSN-18fv. The ejecta structure is represented in the two-dimensional plane with X-axis being the line-of-sight direction and Y being axis perpendicular to that of the plane of sky and line-of-sight. The colour gradient represents the velocity values varying along the line-of-sight. The ejecta structure in the form of grid is over-plotted to show different epochs. A zoomed view of the region highlighted in red in the centre is shown in the \textit{top-right} panel. The inclination angle of the system is the angle between the plane of the sky and orbital plane. \textit{Bottom:} Histogram of the normalised intensity variation along the line-of-sight. The colours represents different epochs. This shows that the asymmetry in the ejecta structure decreases as the system evolves. The intensity distribution along the line-of-sight becomes uniform on to the red and blue components as the system approaches day 316.}
    \label{all_shape}
\end{figure*}

The ejecta geometry of H$\alpha$ (Fig. \ref{all_shape}) on days 18 and 21 are also found to be asymmetric bipolar with an inclination angle of about 53$^\circ$. On day 18 and 21, the red component is 1.7 and 2.2 times bigger than that of the blue component respectively. Also, the red component is 1.3 and 1.2 times brighter than that of the blue component (\textit{bottom panel} in Fig. \ref{all_shape}) on day 18 and 21 respectively.

The ejecta geometry of H$\alpha$ and O I have the extent ratio along the polar axis of 1.11:1 for both the epochs. The similarity in the velocity profiles and the ejecta structures along with the extent of the ejecta of these two lines suggest that they are coming from the same region. It also suggests that the O I line could be excited by H$\alpha$. This is possible when there is optically thick gas in the medium \citep{Str77}.

\subsubsection*{Day 64}
The ejecta geometry of H$\alpha$ on day 64 (2018 May 19) post the re-brightenings continues to be an asymmetric bipolar structure with an inclination angle of about 53$^\circ$. The red component is 1.25 times bigger than that of the blue component along the polar axis (Fig. \ref{all_shape}). The highest peak intensity component is located around the centre which is $\sim$1.6 times higher than that of blue and red components (\textit{bottom panel} in Fig. \ref{all_shape}).

\subsubsection*{Day 96}
The ejecta geometry of H$\alpha$ (Fig. \ref{all_shape}) on this epoch continues to be an asymmetric bipolar structure one with an inclination angle of about 82$^\circ$. The red component is 1.5 times bigger than that of the blue component. However, the highest peak intensity component is located around the centre similar to that of day 64 (\textit{bottom panel} in Fig. \ref{all_shape}). 

\begin{figure}
    \centering
    \includegraphics[width=\columnwidth]{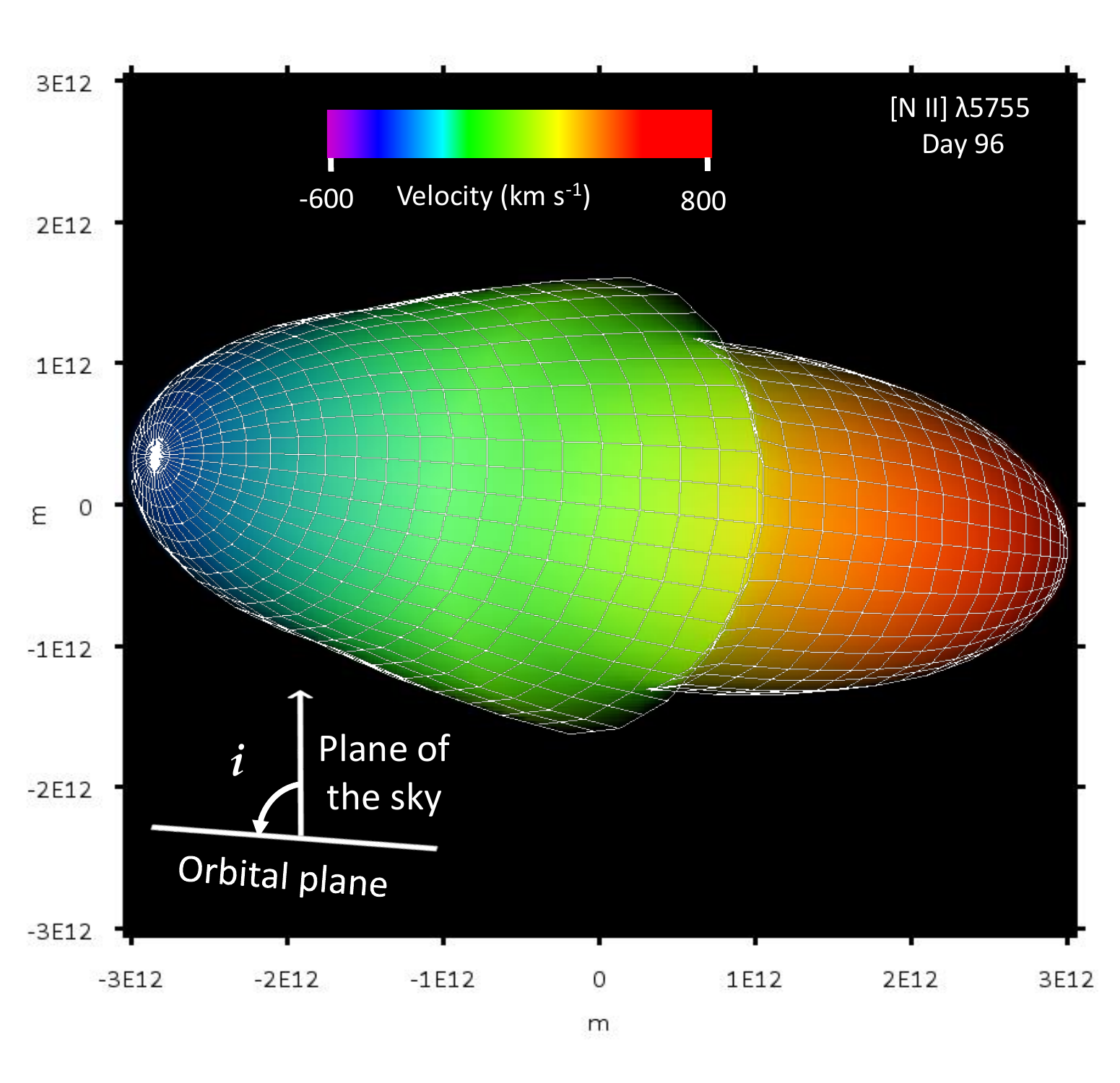}
    \caption{The asymmetric ellipsoidal-like [N II] geometry of the ejecta of ASASSN-18fv in the two-dimensional plane with X-axis being the line-of-sight direction and Y being axis perpendicular to that of the plane of sky and line-of-sight, and color gradient being the velocity values varying along the line-of-sight direction. The structure is obtained using the velocity profile of [N II] 5755 \AA\ obtained on 2018 June 21 (day 96).}
    \label{d96_s}
\end{figure}

The ejecta geometry of [N II] 5755 \AA\ is asymmetric bipolar ellipsoidal-like structure with an inclination angle of about 82$^\circ$. The best-fit [N II] 5755 \AA\ geometrical structure corresponding to its best-fit velocity profile is as shown in Fig. \ref{d96_s}. The blue component is 1.3 times bigger than that of the red component.

The ejecta geometry of H$\alpha$ and [N II] have the extent ratio along the polar axis of 1.88:1. The dissimilarity in the ejecta structures of these two lines and their extents along the polar axis suggest that they are coming from different regions.

\subsubsection*{Day 316}
The velocity line profiles of N III 4638 \AA, He II 4686 \AA, H$\beta$, He I 5876 \AA, H$\alpha$, He I 6678 \AA\ and He I 7065 \AA\ were modelled during this epoch to obtain the ejecta structure. 
The ejecta geometries obtained corresponding to the above mentioned velocity profiles are as follows:
\begin{itemize}
	\item[$\ast$] N III 4638 \AA: Asymmetric bipolar structure with an inclination angle of about 50$^\circ$. The blue component is slightly ($\sim$1.1 times) bigger than that of the red component.
	\item[$\ast$] He II 4686 \AA: Asymmetric bipolar structure with an inclination angle of about 40$^\circ$. The blue component is twice as big and twice the peak intensity value as that of the red component. The polar end of the red component is triangular in shape.
	\item[$\ast$] H$\beta$: Asymmetric bipolar one with an inclination angle of about 50$^\circ$. The blue component is 1.5 times bigger than that of the red component.
	\item[$\ast$] He I 5876 \AA: Asymmetric bipolar one with an inclination angle of about 35$^\circ$. The red component is 1.5 times bigger than that of the blue component. The polar end of the blue component is triangular in shape.
	\item[$\ast$] H$\alpha$: The asymmetry in the H$\alpha$ ejecta geometries seen in the previous epochs shows a decrease trend as the system evolves. On day 316 (2019 Jun 26), the H$\alpha$ geometry is found to be bipolar with equatorial rings with almost symmetric structures with an inclination angle of about 50$^\circ$. The blue component is found to be 1.1 times that of the red component with similar peak intensity values (\textit{bottom panel} in Fig. \ref{all_shape}).
	\item[$\ast$] He I 6678 \AA: Asymmetric bipolar with similar size and intensity values of the blue and red components, with an inclination angle of about 40$^\circ$. The polar end of the blue component is triangular in shape while red component has a more circular polar end.
	\item[$\ast$] He I 7065 \AA: Asymmetric bipolar structure with an inclination angle of about 35$^\circ$. The red component is 1.5 times bigger than that of the blue component. The polar end of the blue component is triangular in shape similar to 6678 \AA\ line.
\end{itemize}
\begin{figure*}
    \centering
    \includegraphics{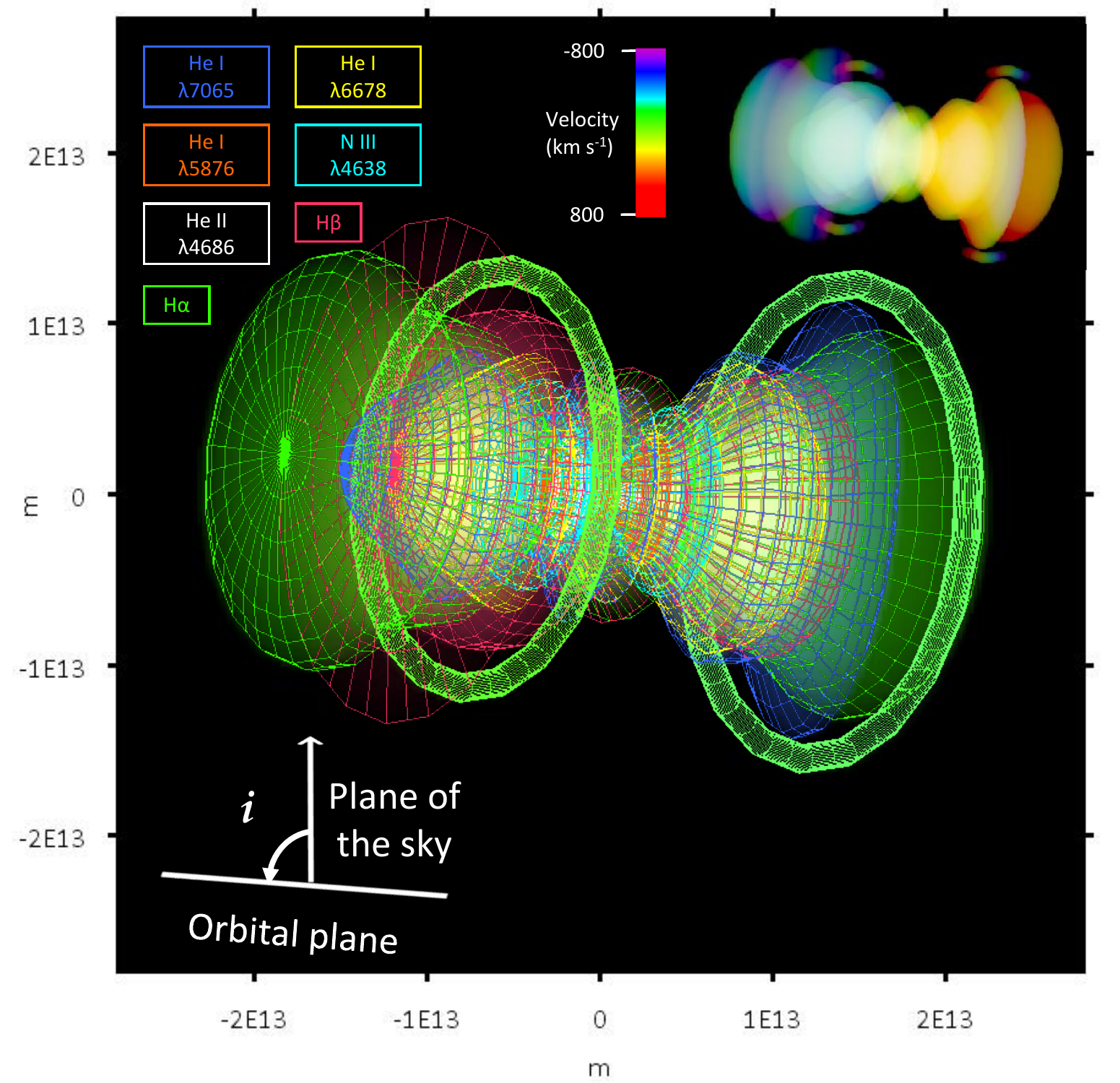}
    \caption{The geometry of the ejecta of ASASSN-18fv obtained on day 316 using H I, He I, He II and N III velocity profiles. The ejecta structure is represented in the two-dimensional plane with the line-of-sight direction along the X-axis and Y axis perpendicular to that of the plane of sky and line-of-sight. The colours representing different line profiles are as indicated. In the \textit{top-right} panel, color gradient represent the velocity values varying along the line-of-sight. H I and N III regions have a more dumbbell-like structure, while the He region is bipolar with triangular polar ends.}
    \label{d316_all}
\end{figure*}
The best-fit geometrical structure of day 316 (2019 Jan 26) corresponding to their best-fit velocity profiles are as shown in Fig. \ref{d316_all}. It is interesting to note that the extent of He II structure along the polar axis is 8 times smaller than that of H$\alpha$ extent suggesting that it is coming from the inner regions of the system. The extent of He I and N III structures are also 2 to 6 times smaller than that of the H$\alpha$ extent. Further, the structures of helium are different from that of N III. This suggests that they are coming from different regions.

The best-fit H$\alpha$ geometrical structure of all the epochs corresponding to their best-fit velocity profiles are as shown in Fig. \ref{all_shape}. An evolution of the ejecta geometry is seen clearly seen along the polar axis. Significant asymmetry of the ejecta is seen in the initial epochs, (zoomed in region in \textit{top-right} panel of Fig. \ref{all_shape}), as also indicated by the emission line profiles. As the system evolves, there is a clear indication of expansion in both polar and equatorial direction. The intensity and velocity variation along the line-of-sight also significant. The geometry of the H$\alpha$ emitting region changes from an asymmetric one to a more symmetric one as the nova evolves to the nebular phase (\textit{bottom} panel of Fig. \ref{all_shape}).

\section{Results and Discussion}
The results obtained from optical spectroscopic observations for the nova ASASSN-18fv are discussed in detail in the previous sections. The distance to the nova and other parameters like absolute magnitude, reddening and mass of the WD are estimated using the AAVSO and SMARTS database. In its early phase of optical spectroscopic evolution, the nova showed low expansion velocity as well as a weak blue continuum quite unlike many classical novae, leading \cite{str18} and \cite{izz18} to suggest the event to be either a young stellar object or a stellar merger event such as luminous red novae or helium-flash explosions. However, a later stage NIR spectrum reported by \cite{rab18} showed the spectrum to be consistent with that of post-maximum classical novae. The optical spectra presented in this study also clearly indicate ASASSN-18fv to be a classical nova explosion. The spectral evolution clearly indicates it to belong to the hybrid class of novae. Coronal lines are absent in the optical spectrum during the late phase. A similar trend of spectral evolution was observed in V5114 Sgr \citep{gca12} which is also a moderately fast nova.

ASASSN-18fv belongs to the speed class of moderately fast novae (t$_2$ $\sim$ 50 days) with small fluctuations in the NIR light curve in early phase followed by a decline in the late phase. In the optical light curve, from day 24--34 re-brightening is seen, followed by a decline. Many novae are known to have re-brightenings during the early phase of their outbursts like V5113 Sgr \citep{kiy04}, V2540 Oph \citep{ak05}, V4745 Sgr \citep{csa05}, V1186 Sco \citep{sch07}, V458 Vul \citep{pog08}, V5558 Sgr \citep{tan11a} and V2676 Oph \citep{Raj17}. Following the re-brightening phases in these novae, the optical spectrum is seen to undergo some changes such as the re-appearance of the P-Cygni profiles \citep{tan11b}. The absorption component that generally disappears during the post maximum decline reappears post the re-brightening. \cite{tan11b} suggest the appearance of the P-Cygni profiles to be due to a re-expansion of the photosphere after it has shifted sufficiently inside. In ASASSN-18fv, the early spectra consists of hydrogen Balmer, Ca II, Na I, O I and Fe II lines similar to a typical Fe II class nova. As the system evolved post the re-brightening, the spectra were dominated by helium and nitrogen lines similar to that of He/N class. The Fe II multiplets slowly faded. The spectrum evolved from Fe II to He/N class of nova.

During the period (day 24--34) when re-brightenings were seen in the optical light curve, a bright and prolonged $\gamma$-ray emission was also reported for the system between day 28 and 33 (2018 April 13 and 18) by \citep{jea18,pia18}. ASASSN-18fv was detected in hard X-rays by NuSTAR during 2018 April 20--22, but not detected in soft X-rays by {\it Swift XRT} \citep{nel18}. Both observations indicate the presence of shock, which could be due to the interaction of white dwarf wind with the initial nova ejection, or due to the interaction of the nova ejecta with a pre-existing material such as wind from the secondary, or a thick circumbinary material. The NuSTAR observations suggest the presence of a deeply embedded internal shock in the ejecta \citep{nel18}, while the presence of a shock heated circumbinary material is suggested by \cite{lou20} based on time-lapsed spectroscopic observations during the early phases. Based on $\gamma$-ray, X-ray, optical and radio observations, \cite{Ayd20} suggest an initial ejection of a slow moving torus followed by the ejection of a faster wind that shock interacts with the earlier ejected material.

The morpho-kinematic H$\alpha$ geometry of the ejecta obtained on day 18 and 21 have asymmetric dumbbell-like structure with brighter high velocity component along the line-of-sight. On day 64 and 96, the H$\alpha$ geometry is more like an intersection of spheres of different radii with high velocity component being bigger and brighter along the line-of-sight. The change in the geometry post day 21 could be due to the internal shocks present in the ejecta as suggested by \cite{nel18} based on the absorbed thermal plasma model which fit the NuSTAR X-ray spectrum or evidence of internal shocks based on $\gamma$-ray emission by \cite{Ayd20}, or the presence of He/N class spectral lines. On day 316, the geometry is back to its dumbbell-like structures, however it is closer to a symmetric morphology. As the shell expands, there is probability of its interaction with the secondary or with any CSM material present. This could lead to the formation of ring-like structures in the ejecta. The geometrical evolution of the system is significant from day 18 to 316 in terms of symmetry. The asymmetry observed in the system gradually decreases with time and the system is almost symmetric on day 316. The decreasing asymmetry is consistent with the behaviour of an expanding shell. The asymmetric ellipsoidal structure of [N II] obtained on day 96 is completely different when compared to that of the structures of O I 8446\AA, H$\alpha$ and He. This suggests that the [N II] is originating from a different region. O I 8446\AA\ region has a geometry and size very similar to that of H$\alpha$ that is, asymmetric dumbbell-like structure suggesting that they could be originating from the same region. The extent of He structures along the polar axis clearly suggest that they are coming from the inner regions of the shell. They have more triangular-like polar ends rather than circular ends seen in other structures.

\section{Summary}
Evolution of the optical spectrum of nova ASASSN-18fv is presented here based on the data obtained from day 6--316 since its discovery. Also presented is the evolution of the morpho-kinematic geometry of the nova ejecta for a few epochs, at different phases. The important results of the analyses are summarised here.
\begin{enumerate}
 \item Using the optical and NIR data from AAVSO and SMARTS, t$_2$ and t$_3$ were estimated to be 50 $\pm$ 5 days and 70 $\pm$ 5 days respectively, indicating ASASSN-18fv belongs to the class of moderately fast novae.
 \item Reddening, \textit{E(B-V)} was estimated to be about 0.75 and distance to the nova was found to be 1.3 kpc.
 \item Mass of the WD was estimated to be 0.7 M$_\odot$.
 \item The optical spectral line profiles evolved from P-Cygni with broad emission and narrow absorption components to emission ones with boxy and structured profiles.
 \item The nova evolved from an Fe II to He/N spectral class. According to the available spectral data, this transition occurred post the period of re-brightening observed in the optical and NIR light curve, and the $\gamma$-ray emission and shock detection period.
 \item The photo-ionization analysis was carried out on day 96. The ejected mass is found to be 6.07 $\times$ 10$^{-4}$ M$_\odot$. The abundance values of nitrogen, oxygen, helium and iron are found to be over abundant compared to solar abundance values.
 \item The H$\alpha$ and O I 8446 \AA\ morphology of the nova ejecta in the early phase are found to be asymmetric bipolar ones with an inclination angle of about 53$^{\circ}$ and 60$^{\circ}$ respectively.
 \item The morpho-kinematic [N II] geometry of the nova ejecta obtained on day 96 is found to be an asymmetric bipolar ellipsoidal-like structure with an inclination angle of about 50$^{\circ}$.
 \item The morpho-kinematic H$\alpha$, H$\beta$ geometries of the nova ejecta in the late phase are found to be an asymmetric bipolar structure along with equatorial rings, with an inclination angle of about 50$^{\circ}$. Presence of equatorial rings in H$\alpha$ geometry suggest the possible interaction of outflow with the secondary. However, the asymmetry seen in this phase is almost negligible with the blue component being just about 1.1 times bigger than that of the red component.
 \item The morpho-kinematic He geometry of the nova ejecta in the late phase is also found to be asymmetric bipolar with an inclination angle of about 37$^{\circ}$, and triangular polar ends.
\end{enumerate}

\section*{Acknowledgements}
The authors would like to thank the referee for critically reading of the manuscript. We acknowledge with thanks the variable star observations from the AAVSO International Database contributed by observers worldwide and used in this research. We also acknowledge the use of SMARTS data. We thank all the observers of VBT and CZT at VBO for accommodating some time for ToO observations. Also, the VBT TAC for the time and support during ToO and regular observations. VBO is operated by the Indian Institute of Astrophysics, Bangalore. AR acknowledges the Research Associate Fellowship with Order No. 03(1428)/18/EMR-II under Council of Scientific and Industrial Research (CSIR).

\noindent \textit{Facilities:} SALT:10m, VBT:2m, CZT:1m.\\
\textit{Software:} IRAF (v2.16.1 \cite{tod93}), Python (v3.6.8), CLOUDY (vC17.01), SHAPE (v5.1), ISIS (v5.9.2 \cite{isis})\\
\textit{Python modules:} pyCloudy (v0.9.8b5), numpy (v1.17.0 \cite{oli06}), scipy (v1.3.1 \cite{Eri01}), matplotlib (v3.1.1 \cite{Hun07}).  

\bibliographystyle{mnras}
\setlength{\bibsep}{0.0pt}
\footnotesize{\bibliography{asa18fv}}

\label{lastpage}
\end{document}